\documentclass{aastex631}

\submitjournal{The Astrophysical Journal}

\shorttitle{Study of the VHE emission of M87 through its broadband SED}
\shortauthors{Alfaro et al.}
\graphicspath{{./}{figures/}}

\begin{document}

\title{Study of the Very High Energy emission of M87 through its broadband spectral energy distribution}

\correspondingauthor{Fernando Ure\~{n}a-Mena \\ furena@inaoep.mx}
\correspondingauthor{Daniel Rosa-Gonz\'{a}lez\\ danrosa@inaoep.mx}
\correspondingauthor{Alberto Carrami\~{n}ana \\ alberto@inaoep.mx}
\correspondingauthor{Anna Lia Longinotti \\ alonginotti@astro.unam.mx }

\author[0000-0001-8749-1647]{R.~Alfaro}
\affiliation{Instituto de F\'{i}sica, Universidad Nacional Aut\'{o}noma de M\'{e}xico, Ciudad de Mexico, Mexico }

\author{C.~Alvarez}
\affiliation{Universidad Aut\'{o}noma de Chiapas, Tuxtla Guti\'{e}rrez, Chiapas, M\'{e}xico}

\author{J.C.~Arteaga-Vel\'{a}zquez}
\affiliation{Universidad Michoacana de San Nicol\'{a}s de Hidalgo, Morelia, Mexico }

\author[0000-0002-4020-4142]{D.~Avila Rojas}
\affiliation{Instituto de F\'{i}sica, Universidad Nacional Aut\'{o}noma de M\'{e}xico, Ciudad de Mexico, Mexico }

\author[0000-0002-2084-5049]{H.A.~Ayala Solares}
\affiliation{Department of Physics, Pennsylvania State University, University Park, PA, USA }

\author[0000-0003-3207-105X]{E.~Belmont-Moreno}
\affiliation{Instituto de F\'{i}sica, Universidad Nacional Aut\'{o}noma de M\'{e}xico, Ciudad de Mexico, Mexico }

\author[0000-0003-2158-2292]{T.~Capistr\'{a}n}
\affiliation{Instituto de Astronom\'{i}a, Universidad Nacional Aut\'{o}noma de M\'{e}xico, Ciudad de Mexico, Mexico }

\author[0000-0002-8553-3302]{A.~Carrami\~{n}ana}
\affiliation{Instituto Nacional de Astrof\'{i}sica, \'{o}ptica y Electr\'{o}nica, Puebla, Mexico }

\author[0000-0002-6144-9122]{S.~Casanova}
\affiliation{Institute of Nuclear Physics Polish Academy of Sciences, PL-31342 IFJ-PAN, Krakow, Poland }

\author[0000-0002-7607-9582]{U.~Cotti}
\affiliation{Universidad Michoacana de San Nicol\'{a}s de Hidalgo, Morelia, Mexico }

\author[0000-0002-1132-871X]{J.~Cotzomi}
\affiliation{Facultad de Ciencias F\'{i}sico Matem\'{a}ticas, Benem\'{e}rita Universidad Aut\'{o}noma de Puebla, Puebla, Mexico }

\author[0000-0002-7747-754X]{S.~Couti\~{n}o de Le\'{o}n}
\affiliation{Department of Physics, University of Wisconsin-Madison, Madison, WI, USA }

\author[0000-0001-9643-4134]{E.~De la Fuente}
\affiliation{Departamento de F\'{i}sica, Centro Universitario de Ciencias Exactas e Ingenierias, Universidad de Guadalajara, Guadalajara, Mexico }

\author[0000-0002-8528-9573]{C.~de Le\'{o}n}
\affiliation{Universidad Michoacana de San Nicol\'{a}s de Hidalgo, Morelia, Mexico }

\author{R.~Diaz Hernandez}
\affiliation{Instituto Nacional de Astrof\'{i}sica, \'{o}ptica y Electr\'{o}nica, Puebla, Mexico }

\author[0000-0002-2987-9691]{M.A.~DuVernois}
\affiliation{Department of Physics, University of Wisconsin-Madison, Madison, WI, USA }

\author[0000-0003-2169-0306]{M.~Durocher}
\affiliation{Physics Division, Los Alamos National Laboratory, Los Alamos, NM, USA }

\author[0000-0002-0087-0693]{J.C.~D\'{i}az-V\'{e}lez}
\affiliation{Departamento de F\'{i}sica, Centro Universitario de Ciencias Exactas e Ingenierias, Universidad de Guadalajara, Guadalajara, Mexico }

\author[0000-0001-7074-1726]{C.~Espinoza}
\affiliation{Instituto de F\'{i}sica, Universidad Nacional Aut\'{o}noma de M\'{e}xico, Ciudad de Mexico, Mexico }

\author{K.L.~Fan}
\affiliation{Department of Physics, University of Maryland, College Park, MD, USA }

\author{M.~Fern\'{a}ndez Alonso}
\affiliation{Department of Physics, Pennsylvania State University, University Park, PA, USA }

\author[0000-0002-0173-6453]{N.~Fraija}
\affiliation{Instituto de Astronom\'{i}a, Universidad Nacional Aut\'{o}noma de M\'{e}xico, Ciudad de Mexico, Mexico }

\author[0000-0002-4188-5584]{J.A.~Garc\'{i}a-Gonz\'{a}lez}
\affiliation{Tecnologico de Monterrey, Escuela de Ingenier\'{i}a y Ciencias, Ave. Eugenio Garza Sada 2501, Monterrey, N.L., Mexico, 64849}

\author[0000-0003-1122-4168]{F.~Garfias}
\affiliation{Instituto de Astronom\'{i}a, Universidad Nacional Aut\'{o}noma de M\'{e}xico, Ciudad de Mexico, Mexico }

\author[0000-0002-5209-5641]{M.M.~Gonz\'{a}lez}
\affiliation{Instituto de Astronom\'{i}a, Universidad Nacional Aut\'{o}noma de M\'{e}xico, Ciudad de Mexico, Mexico }

\author[0000-0002-9790-1299]{J.A.~Goodman}
\affiliation{Department of Physics, University of Maryland, College Park, MD, USA }

\author[0000-0001-9844-2648]{J.P.~Harding}
\affiliation{Physics Division, Los Alamos National Laboratory, Los Alamos, NM, USA }

\author[0000-0002-3808-4639]{D.~Huang}
\affiliation{Department of Physics, Michigan Technological University, Houghton, MI, USA }

\author[0000-0002-5527-7141]{F.~Hueyotl-Zahuantitla}
\affiliation{Universidad Aut\'{o}noma de Chiapas, Tuxtla Guti\'{e}rrez, Chiapas, M\'{e}xico}

\author{P.~Hüntemeyer}
\affiliation{Department of Physics, Michigan Technological University, Houghton, MI, USA }

\author[0000-0003-4467-3621]{V.~Joshi}
\affiliation{Erlangen Centre for Astroparticle Physics, Friedrich-Alexander-Universit\"at Erlangen-N\"urnberg, Erlangen, Germany}

\author[0000-0001-5516-4975]{H.~Le\'{o}n Vargas}
\affiliation{Instituto de F\'{i}sica, Universidad Nacional Aut\'{o}noma de M\'{e}xico, Ciudad de Mexico, Mexico }

\author[0000-0003-2696-947X]{J.T.~Linnemann}
\affiliation{Department of Physics and Astronomy, Michigan State University, East Lansing, MI, USA }

\author[0000-0001-8825-3624]{A.L.~Longinotti}
\affiliation{Instituto de Astronom\'{i}a, Universidad Nacional Aut\'{o}noma de M\'{e}xico, Ciudad de Mexico, Mexico }

\author[0000-0003-2810-4867]{G.~Luis-Raya}
\affiliation{Universidad Politecnica de Pachuca, Pachuca, Hgo, Mexico }

\author[0000-0001-8088-400X]{K.~Malone}
\affiliation{Physics Division, Los Alamos National Laboratory, Los Alamos, NM, USA }

\author[0000-0001-9052-856X]{O.~Martinez}
\affiliation{Facultad de Ciencias F\'{i}sico Matem\'{a}ticas, Benem\'{e}rita Universidad Aut\'{o}noma de Puebla, Puebla, Mexico }

\author[0000-0002-2824-3544]{J.~Mart\'{i}nez-Castro}
\affiliation{Centro de Investigaci\'on en Computaci\'on, Instituto Polit\'ecnico Nacional, M\'exico City, M\'exico.}

\author[0000-0002-2610-863X]{J.A.~Matthews}
\affiliation{Dept of Physics and Astronomy, University of New Mexico, Albuquerque, NM, USA }

\author[0000-0002-8390-9011]{P.~Miranda-Romagnoli}
\affiliation{Universidad Aut\'{o}noma del Estado de Hidalgo, Pachuca, Mexico }

\author[0000-0002-1114-2640]{E.~Moreno}
\affiliation{Facultad de Ciencias F\'{i}sico Matem\'{a}ticas, Benem\'{e}rita Universidad Aut\'{o}noma de Puebla, Puebla, Mexico }

\author[0000-0002-7675-4656]{M.~Mostaf\'{a}}
\affiliation{Department of Physics, Pennsylvania State University, University Park, PA, USA }

\author[0000-0003-0587-4324]{A.~Nayerhoda}
\affiliation{Institute of Nuclear Physics Polish Academy of Sciences, PL-31342 IFJ-PAN, Krakow, Poland }

\author[0000-0003-1059-8731]{L.~Nellen}
\affiliation{Instituto de Ciencias Nucleares, Universidad Nacional Aut\'{o}noma de Mexico, Ciudad de Mexico, Mexico }

\author[0000-0001-7099-108X]{R.~Noriega-Papaqui}
\affiliation{Universidad Aut\'{o}noma del Estado de Hidalgo, Pachuca, Mexico }

\author[0000-0002-5448-7577]{N.~Omodei}
\affiliation{Department of Physics, Stanford University: Stanford, CA 94305–4060, USA}

\author{A.~Peisker}
\affiliation{Department of Physics and Astronomy, Michigan State University, East Lansing, MI, USA }

\author[0000-0001-5998-4938]{E.G.~P\'{e}rez-P\'{e}rez}
\affiliation{Universidad Politecnica de Pachuca, Pachuca, Hgo, Mexico }

\author[0000-0002-6524-9769]{C.D.~Rho}
\affiliation{University of Seoul, Seoul, Rep. of Korea}

\author[0000-0003-1327-0838]{D.~Rosa-Gonz\'{a}lez}
\affiliation{Instituto Nacional de Astrof\'{i}sica, \'{o}ptica y Electr\'{o}nica, Puebla, Mexico }

\author{H.~Salazar}
\affiliation{Facultad de Ciencias F\'{i}sico Matem\'{a}ticas, Benem\'{e}rita Universidad Aut\'{o}noma de Puebla, Puebla, Mexico }

\author{D.~Salazar-Gallegos}
\affiliation{Department of Physics and Astronomy, Michigan State University, East Lansing, MI, USA }

\author[0000-0002-8610-8703]{F.~Salesa Greus}
\affiliation{Institute of Nuclear Physics Polish Academy of Sciences, PL-31342 IFJ-PAN, Krakow, Poland }

\author[0000-0001-6079-2722]{A.~Sandoval}
\affiliation{Instituto de F\'{i}sica, Universidad Nacional Aut\'{o}noma de M\'{e}xico, Ciudad de Mexico, Mexico }

\author{J.~Serna-Franco}
\affiliation{Instituto de F\'{i}sica, Universidad Nacional Aut\'{o}noma de M\'{e}xico, Ciudad de Mexico, Mexico }

\author{Y.~Son}
\affiliation{University of Seoul, Seoul, Rep. of Korea}

\author[0000-0002-1492-0380]{R.W.~Springer}
\affiliation{Department of Physics and Astronomy, University of Utah, Salt Lake City, UT, USA }

\author{O.~Tibolla}
\affiliation{Universidad Politecnica de Pachuca, Pachuca, Hgo, Mexico }

\author[0000-0001-9725-1479]{K.~Tollefson}
\affiliation{Department of Physics and Astronomy, Michigan State University, East Lansing, MI, USA }

\author[0000-0002-1689-3945]{I.~Torres}
\affiliation{Instituto Nacional de Astrof\'{i}sica, \'{o}ptica y Electr\'{o}nica, Puebla, Mexico }

\author[0000-0002-2748-2527]{F.~Ure\~{n}a-Mena}
\affiliation{Instituto Nacional de Astrof\'{i}sica, \'{o}ptica y Electr\'{o}nica, Puebla, Mexico }

\author[0000-0001-6876-2800]{L.~Villase\~{n}or}
\affiliation{Facultad de Ciencias F\'{i}sico Matem\'{a}ticas, Benem\'{e}rita Universidad Aut\'{o}noma de Puebla, Puebla, Mexico }

\author{X.~Wang}
\affiliation{Department of Physics, Michigan Technological University, Houghton, MI, USA }

\author[0000-0002-6623-0277]{E.~Willox}
\affiliation{Department of Physics, University of Maryland, College Park, MD, USA }

\author[0000-0001-9976-2387]{A.~Zepeda}
\affiliation{Physics Department, Centro de Investigacion y de Estudios Avanzados del IPN, Mexico City, Mexico}

\collaboration{100}{HAWC Collaboration}



\begin{abstract}

The radio galaxy M87 is the central dominant galaxy of the Virgo Cluster.  Very High Energy  (VHE,$\gtrsim 0.1$ TeV)  emission, from M87 has been detected by Imaging Air Cherenkov Telescopes (IACTs ). Recently,  marginal evidence for VHE  long-term  emission  has also been observed  by the High Altitude Water  Cherenkov (HAWC) Observatory, a gamma ray and cosmic-ray
detector array located in Puebla, Mexico.  The mechanism that produces VHE emission in M87 remains unclear. This emission is originated in its prominent jet, which has been spatially resolved from radio to X-rays. 
In this paper, we constructed a spectral energy distribution from radio to 
gamma rays that  is representative of the non-flaring activity of the source, and in order to explain the observed emission, we fit it with a lepto-hadronic emission model.  
We found that this model is able to explain non-flaring VHE emission of M87 as well as an orphan flare reported in 2005.  

\end{abstract}

\keywords{Active galactic nuclei; Gamma rays; Gamma ray sources; Radio galaxies }


\section{Introduction} \label{sec:intro}

Gamma rays constitute the highest energy electromagnetic radiation tracing the most energetic phenomena in the Universe.  Active Galactic Nuclei (AGN) are  important sources of extragalactic gamma rays, and according to the current consensus they are powered by accreting super massive black holes (SMBH). Most  AGNs that are VHE emitters are classified as blazars, i.e. radio-loud AGNs whose jets are pointing nearly towards the observer. Since particles within the jet travel at nearly the speed of light, relativistic beaming increases the brightness of these objects.  
According to unification schemes \citep{urry1995unified},  radio galaxies (RDGs) correspond to the misaligned counterparts of blazars, and so far, VHE emission has been detected from six of them \citep{rieger2018radio}.
Since RDGs are located on average closer  than blazars, it is possible obtain more detailed observations to test theoretical models of their emission. The broadband spectral energy distributions (SED) of AGN jets, which are prominent in the emission of blazars and RDGs, are globally non thermal and they are characterized by the presence of two components (generally referred to as ``peaks") \citep{blandford2019relativistic}. The low energy component is usually attributed to synchrotron emission, which is produced when relativistic particles are moving in the presence of a magnetic field \citep{rybicki2008radiative}. On the other hand,  the second peak has been explained by many different models. These models can be divided into two types: leptonic, where the high energy component of the broadband SED is explained as inverse Compton (IC) emission produced  by an electron population in the jet \citep{longair2011high},  and hadronic, where the second component  is produced by mechanisms involving the collision of accelerated protons with the surrounding environment \citep{mucke2003bl}. \\

M87 (R.A. $12^{\text{h}}30^{\text{m}} 47^{\text{s}}.2 $ Dec.$+12^\circ 23^{\prime} 51 ^{\prime\prime}$), which is classified as a giant Fanaroff-Riley I (FR-I) RDG,  is the  central dominant galaxy of the Virgo Cluster. It is an elliptical galaxy with a diameter of $\sim$300 kpc \citep{doherty2009edge}, a dynamical mass within 180 kpc estimated as $(1.5\pm0.2)\times10^{13} \ M_\odot$ \citep{zhu2014next} and a redshift of $z=0.0044$. It is located at a distance of 16.4 $\pm$ 0.5 Mpc, which is a redshift independent measurement \citep{bird2010inner}. M87 hosts a SMBH, named M87*, whose shadow was the first imaged by the Event Horizon Telescope \citep{event2019firstI}. The prominent jet of M87 is one its most noticeable characteristics. This jet has been studied for the last one hundred years \citep{curtis1918descriptions}, it has a length of about 2.5 kpc \citep{biretta1999hubble}  and  has been resolved from radio to X-rays. It presents complex structures like knots and diffuse emission \citep{perlman1999optical,perlman2001deep}, apparent superluminal motion \citep{cheung2007superluminal} and a complex variability \citep{harris2006outburst}.\\

M87 was the first RDG detected at VHE \citep{aharonian2003giant}.  It has been detected by different IACTs  such as HESS \citep{Aharonian2006fast}, VERITAS \citep{acciari2010veritas} and MAGIC. \citep{acciari2020monitoring}
 Recently, the HAWC Collaboration \citep{agnsurvey} reported weak evidence (3.6$\sigma$) of long-term VHE emission from this source.  M87 has shown a complex behavior at VHE \citep{benkhali2019complex} with a rapid variability during flaring states \citep{abramowski20122010}. According to available observations, three VHE flares from M87 have been detected in 2005  \citep{Aharonian2006fast}, 2008 \citep{albert2008very} and 2010 \citep{abramowski20122010}. Gamma ray angular resolution is not sufficient to determine the region of the galaxy where this emission is produced. Variability studies have suggested that the innermost jet zone of M87 is  most likely  the source of its VHE emission. The other candidate location is the jet feature  \textit{HST}-1, but it is disfavored by the VHE timescale variability and the lack of correlated activity in other bands during TeV flares \citep{abramowski20122010,benkhali2019complex}.  \\
 
 The broadband emission of M87 has usually been explained by a one-zone Synchrotron Self Compton (SSC)  scenario \citep{abdo2009fermi,de2015high}. In these models, the first component is attributed to synchrotron radiation that is produced by   electrons moving at relativistic velocity with random orientation  with respect to the magnetic  field. The second peak is explained by inverse Compton scattering of synchrotron photons to higher energies by the same electron population. However, some authors claim that SSC models are not able to explain VHE emission in M87. Evidence for a possible spectral turnover in the GeV regime $E\gtrsim 10$ GeV was found by \cite{benkhali2019complex}. This was interpreted as due to the presence of an additional physical component in the emission. However, constraining the decrease of this component from only \textit{Fermi} data is not possible and long-term TeV observations are needed.  One zone SSC models were  found to have difficulties in explaining VHE/X-ray correlated variability in M87 \citep{abramowski20122010}, and \cite{fraija2016neutrino}  claimed that  one-zone SSC models cannot be extended to VHE in FR-I RDGs. This is why  alternative ideas have been proposed to explain the SED such as seed photons coming from other regions in the jet \citep{georganopoulos2005core} and  photo-hadronic interactions \citep{fraija2016neutrino}.\\

 The main goal of this work is comparing the VHE emission of the RDG M87 observed by IACTs  during specific epochs  (including the 2005 flare) with the long term quiescent/average emission provided by continuous observation by the HAWC observatory from 2014 to 2019. We used a lepto-hadronic model, which combines SSC and  photo-hadronic scenarios, to explain this emission. We developed a Python  code to simulate the broadband emission of M87 and constructed an average SED of M87 collecting multifrequency observations from data archives. Preliminary results of this work were released in \cite{urena2021}.\\

  \section{Data}

We collected historical archive data  to construct the broadband SED of M87.   Data sets from radio to X-rays were partially based on those  observations of the innermost jet zone used by \cite{abdo2009fermi}, \cite{fraija2016neutrino} and \cite{prieto2016central}. MeV-GeV gamma ray data were obtained from the \textit{Fermi} Large Area Telescope Fourth Source Catalog (4FGL), which is based on the first eight years of data from  the \textit{Fermi} Gamma-ray Space Telescope \citep{fermi2019fermi}\footnote{\url{https://fermi.gsfc.nasa.gov/ssc/data/access/lat/8yr_catalog/}}. The 4FGL covers an energy range  from 50 MeV to 1 TeV.  Four different sets of data were used for the TeV range: 1)  H.E.S.S observations from 2004, which were taken during a TeV quiescent phase \citep{Aharonian2006fast}. 2) H.E.S.S observations from 2005, which were taken during a TeV high activity state but  without evidence of inner jet activity in the rest of the broadband inner jet  spectrum \citep{Aharonian2006fast}. That is why we use  the same broadband SED as in the  non-flaring state case. 3) MAGIC-I observations from 2005-2007, which correspond to an observation campaign where no flaring activity was detected \citep{aleksic2012magic}. 4)  HAWC observations, which cover a quiescent period from 2014 to 2019 corresponding to 1,523 days. \citep{agnsurvey}.\\
  
  The HAWC array consists of 300 water Cherenkov detectors (WCD), each with 4 photomultiplier tubes (PMT). The HAWC data is divided into nine bins according to the fraction of channels hit, which are used to estimate the energy of the events (see \cite{abeysekara2017observation} for more details). In \cite{agnsurvey} a power law spectrum, with a spectral index set to 2.5, was fit to a sample of 138 nearby AGN. For those sources with a $TS>9$, including M87,  an optimized spectrum with free normalization and spectral index was obtained. Then, quasi differential flux limits were obtained in three bands: $0.5-2.0$ TeV, $2.0-8.0$ TeV and $8.0-32.0$ TeV.

\section{Emission model} \label{sec:emission}

We used a hybrid model in this work. Emission components from radio to GeV gamma rays are explained with an SSC scenario whereas photo-hadronic interactions are added to explain the VHE emission. Therefore, the broadband SED has been modeled with three components.  Due to the low redshift of the source, extragalactic background light (EBL) absorption was not relevant for photon energies $\lesssim 10 $ TeV and we did not consider it in the modeling. HAWC data, which are the only data set affected by some relevant EBL absorption, were already corrected for this effect \citep{agnsurvey} using the EBL model of \cite{dominguez2011extragalactic}.   We used the one-zone SSC code described in \cite{finke2008synchrotron}, which considers a homogeneous spherical region or blob in the inner jet moving with a Lorentz factor $\Gamma$  and a randomly oriented magnetic field with mean intensity $B$.  The Doppler factor $\delta$ is given by: \\ 
\begin{equation}
    \delta=[\Gamma(1-\beta\mu)]^{-1},
    \label{eq:doppler}
\end{equation}
where $\beta$ is the ratio of the speed of the jet and the speed of light and $\mu=\cos\theta$ where $\theta$ is the angle of the jet with the observer’s
line of sight.

The minimum variability timescale is:
\begin{equation}
t_{v,min}=\frac{(1+z)R_b^{\prime}}{c\delta },
\label{eq:tvmin}
\end{equation}
where $R_b^\prime$ corresponds to the comoving radius of the region, $c$ to the speed of light, $\delta$ to the Doppler factor and $z$ to the redshift of the source. Comoving quantities are primed following the  convention used by \cite{finke2008synchrotron}.\\

The electron population  of the region, which follows an energy distribution $N^\prime(\gamma^\prime)$, is moving in a randomly oriented magnetic field producing synchrotron radiation. The electron energy  distribution for this model was assumed to be a broken power law given by,

\begin{equation}
     N_e(\gamma^{\prime})=K_e\scalebox{2}{$\bigg\{$}^{\scalebox{1}{$\left(\frac{\gamma^\prime}{\gamma_{break}^\prime}\right)^{-p_1} $ for $\gamma^\prime_1<\gamma^\prime<\gamma^\prime_{break}$}}_{\scalebox{1}{$\left(\frac{\gamma^\prime}{\gamma^{\prime}_{break}}\right)^{-p_2}$for $\gamma^\prime_{break}<\gamma^\prime<\gamma^\prime_2$}},
     \label{eq:Ne}
\end{equation}
where $\gamma^\prime$ is the electron Lorentz factor, $p_1$ and $p_2$ are the power law indices, $\gamma^\prime_{break}$ is the break electron Lorentz factor, $K_e$ is a normalization constant,$\gamma^\prime_1$ and $\gamma_2^\prime$ are the minimum and maximum electron Lorentz factor. 

The photo-hadronic component is based on the model presented by \cite{sahu2019multi2}. Recent evidence of neutrino emission from AGNs seems to support the relevance of photo-hadronic interactions \citep{aartsen2018neutrino}. An accelerated proton population is assumed to be contained in a spherical volume of radius R$^\prime_f$  inside the blob of radius R$^\prime$ (SSC blob) with R$^\prime_f<$R$^\prime$. The inner region is also assumed to have a higher seed photon density because the low density in the SSC emission zone makes the photo-hadronic process inefficient.  The proton population has a power law energy distribution \citep{fraija2016neutrino,sahu2019multi}:
\begin{equation}
\frac{\mathrm{d} N_p}{\mathrm{d} E_p}\propto E_p^{-\alpha}, 
\end{equation}
where the spectral index is $\alpha>2$.

 Due to the higher photon density in this inner volume, protons interact with the background photons through the following mechanism \citep{dermer2009high}: 
\begin{equation}
p+\gamma\to\Delta^{+}\to \bigg\{ ^{\scalebox{1}{ $p + \pi^0 \to p+2\gamma$ } }_{\scalebox{1}{$n+\pi^+\to n +e+3\nu \to p+2e+4\nu$ }}.
\end{equation}

 
 This process requires the center of mass energy of the interaction to exceed the $\Delta$-mass \citep{sahu2019multi2,fraija2016neutrino},
\begin{equation}
E_p^{\prime}\epsilon^{\prime}_\gamma=\frac{(m_\Delta^2-m^2_p)}{2(1-\beta_p\cos\theta)}\cong 0.32\mathrm{GeV^2},
\end{equation}
where $E_p$ is the energy of the proton and $\epsilon_\gamma$ is the energy of the target photon. Considering collisions with SSC photons from all directions, $\beta_p\approx 1$ and viewing from the observer frame:
\begin{equation}
\epsilon_\Gamma\epsilon_\gamma\cong 0.32\frac{\delta^2}{(1+z)^2}\mathrm{GeV^2},
\end{equation}
where $\epsilon_\Gamma$ is the energy of the emitted photon. According to \cite{sahu2019multi2} the $\pi^0$ decay photon flux is given by:
\begin{equation}
f^{p\gamma}(\epsilon_\Gamma)={A_\gamma} f^{ICS} (\nu_{\gamma}) \left(\frac{\epsilon_\Gamma}{\text{TeV}}\right)^{{-\alpha}+3},
\label{eq:ph}
\end{equation}
where $\nu_{\gamma}$ is the frequency of a photon with energy $\epsilon_{\gamma}$ and  $f^{ICS}(\nu_\gamma)$ is the flux at  $\nu_\gamma$. Thus, the total emitted flux at VHE energies is given by the sum of the inverse Compton flux and photo-hadronic flux:
\begin{equation}
f^{VHE}(\nu_\Gamma)=f^{ICS}(\nu_\Gamma)+f^{p\gamma}(\epsilon_\Gamma(\nu_\Gamma)),
\end{equation}
where is $\nu_\Gamma$ is the frequency of the emitted photon and $\epsilon_\Gamma$ the energy of the emitted photon ($\epsilon_\Gamma=h\nu_\Gamma$ where $h$ is the Planck constant). 

The contribution of proton-proton interactions would require a very proton-loaded jet to be significant \citep{reynoso2011lepto}, which is why we did not consider it in this work. Synchrotron emission of protons and muons would need a much higher magnetic field intensity to be relevant. We also neglected synchrotron of pions  because of their short lifetime. 

\section{Methodology}
 We developed a Python code to reproduce the lepto-hadronic model described above. The broadband SED was fit with this emission model.  We obtained the best fit values for the physical parameters and estimated their errors using Monte Carlo simulations.

\subsection{Fitting Technique: SSC model} \label{subsec:SSCfitl}

According to \cite{finke2008synchrotron}, the model shows low dependence on the minimum and maximum electron Lorentz factors  ($\gamma_1^\prime$ and $\gamma_2^\prime$ respectively). That is why they were fixed to the values given by \cite{abdo2009fermi}  $\gamma_1^\prime=1$  and  $\gamma_2^\prime=10^7$. The minimum variability timescale  was assumed to be $t_
{v,min}=1.2\times10^{5}$ s$=1.4$ days, which according to Equation \ref{eq:tvmin} with $\delta=3.9$ corresponds to an emission zone radius of $R_b^{\prime}=1.4\times10
^{16} \ \text{cm} = 4.5  \ \text{mpc}$   chosen by \cite{abdo2009fermi} for being consistent with the highest resolution of VLBA observations and the few day timescale TeV variability. Due to its high degeneracy with $B$ and $\delta$ \citep{yamada2020variations}, the electron spectral normalization constant  was fixed to $K_e^\prime=10^{46}$ \citep{abdo2009fermi}.   Therefore, we used five fitting parameters in the SSC model: mean magnetic field intensity ($B$),  Doppler factor ($\delta$) and the electron energy distribution parameters, the power law indices ($p_1$ and $p_2$) and the break Lorentz factor ( $\gamma_c^\prime$).\\

The fitting technique was based on the method used by \cite{finke2008synchrotron}. We used the results obtained by \cite{abdo2009fermi} as initial values for the fitting parameters: magnetic field $B=0.055$ G, Doppler factor $\delta=3.9$, electron distribution power law index for low energies  $p_1=1.6$, electron distribution power law index for high energies $p_2=3.6$ and electron distribution cutoff  Lorentz factor $\gamma_c^\prime=4000$.  First, the values of $B$ and $\delta$ were fixed  while the electron distribution parameters ($p_1$,$p_2$,$\gamma_c$) were varied in a set of  quantities centered on the initial values.  $B$ and $\delta$  are not correlated with  the electron distribution parameters \citep{yamada2020variations}. The SSC SED was calculated for each combination of generated values and the $\chi^2$ with the observed  data (without including TeV measurements) was obtained for each of them. Because of the importance of the X-ray data to explain the gamma ray fluxes, we excluded those solutions that exceed the \textit{Swift}/BAT upper limits.  The set with the minimum $\chi^2$  was defined as the new  set of initial values. Then, the process was iteratively  repeated until it converged. After finding the best values for the electron distribution parameters, they were kept constant while $B$ and $\delta$ were varied. Following the same procedure, we obtained the best fit values for those other two parameters.\\

Uncertainties were estimated using Monte Carlo simulations  as explained in \cite{press2007numerical}. We draw 10 000  random values for each  observed data point from their  error distributions using the Python function $numpy.random.normal$ \footnote{\url{https://numpy.org/doc/stable/reference/random/generated/numpy.random.normal.html?numpy.random.normal}}. The error distributions were assumed to be normal and centered  in their observed fluxes with their reported errors as standard deviations.  The same number (10 000) of synthetic SEDs were created using the  random values. The best  fit values for each synthetic SED were obtained using the procedure described above with the best fit values  for the observed data as initial values. The error distributions were made using the $\mathrm{10 \ 000}$ best fit values for each parameter. 

\subsection{Fitting Technique: Photo-hadronic component}
\label{subsec:phfit}

We separately fit each set of data with the photo-hadronic model. As some authors report a spectral turnover at $\sim$ 10 GeV \citep{benkhali2019complex}, the two last \textit{Fermi}-LAT data points were also used in this fit.  The corresponding value of $f^{ssc}(\nu_\gamma)$  was calculated for each observed data point   with the   best fit values already obtained  for the SSC model parameters.   We defined a set of possible values for $\alpha$ and $A_\gamma$. The  gamma ray flux was calculated for each combination of parameters at each gamma ray frequency (including both TeV and MeV-GeV data). The gamma ray flux was calculated as the sum of the flux corresponding to the photo-hadronic component and the  already obtained inverse Compton component.  The $\chi^2$ with the observed data fluxes were calculated for each combination of $\alpha$ and $A_\gamma$ and the parameters with the minimum $\chi^2$ were defined as the best fit values. \\

The procedure to estimate errors in the photo-hadronic component was similar to the method  used for the SSC component. It was necessary to generate random samples of values for the SSC model parameters. They were drawn from normal distributions centered on their previously obtained best fit values with their errors as standard deviations. After generating 10 000 random values for each parameter, the same number of random values for the VHE fluxes were  generated. They were drawn from normal distributions centered on the observed fluxes and  with the observational errors as standard deviations.  We used the Python function $numpy.random.normal$ in both cases. Then, 10 000  VHE synthetic SEDs were constructed using the random values  for the TeV fluxes. These SEDs were fit to the photo-hadronic component model without fixing the SSC model parameters but using the 10 000 random values generated for them. We did this to propagate the errors obtained in the fit of the other two components. Each of the 10 000 SSC parameter random samples correlated with one of the 10 000 VHE samples. The best fit values  for each synthetic VHE SED were obtained using the procedure described in Section \ref{subsec:phfit}.  Using all those results, we obtained error distributions for the two model parameters. 

\section{Results}

\begin{figure}[ht!]
    \centering
    \plotone{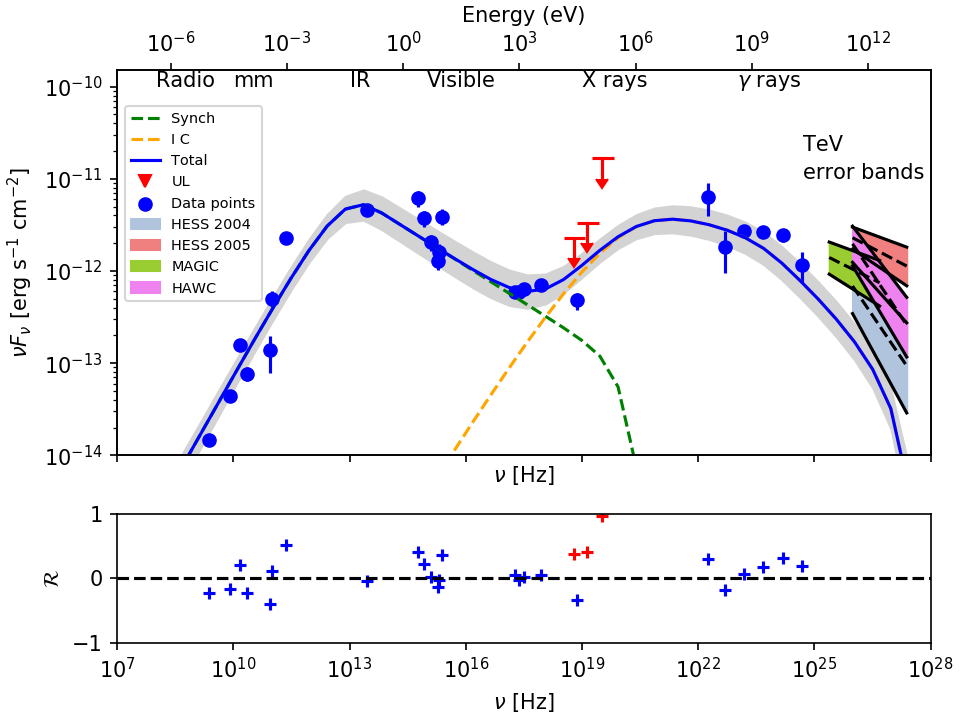}
    \caption{SED of M87 with the best fit SSC model. Blue points correspond to measured  fluxes taken from \cite{morabito1986vlbi}, \cite{morabito1988evidence}, \cite{junor1995radio}, \cite{lee2008global}, \cite{lonsdale19983}, \cite{doeleman2012jet}, \cite{biretta1991radio}, \cite{perlman2001deep}, \cite{sparks1996jet}, \cite{marshall2002high}, \cite{wong2017hard}, \cite{abdo2009fermi} and the 4FGL catalog \cite{fermi2019fermi}.  The model of the synchrotron component is  the orange dashed curve  and the model of the inverse Compton component is the green dashed curve. \textit{Swift}/BAT upper limits obtained by \cite{abdo2009fermi} are shown by red triangles. The gray region corresponds to the 1 $\sigma$ error of the best fit model parameters. For comparison, TeV error bands from 2004 H.E.S.S (blue) \citep{Aharonian2006fast}, 2005 H.E.S.S (red) \citep{Aharonian2006fast}, MAGIC (green) \citep{aleksic2012magic}  and HAWC (violet) \citep{agnsurvey} are shown.  Residuals of the best fit model, which are defined as $\mathcal{R}=\log(F_{\nu,obs}/F_{\nu,mod})$ where  $F_{\nu,obs}$ and $F_{\nu,mod}$ are the observed and predicted flux respectively, are shown in the bottom panel.}
    \label{fig:resSSC}
\end{figure}

 We carried out the SSC model fitting  following the procedure described  in Section \ref{subsec:SSCfitl}. The best fit values of the physical parameters are presented in Table \ref{tab:resultsSSCwitherrors}. The best fit model is plotted together with the observational data and residuals in Figure \ref{fig:resSSC}. 

\begin{deluxetable*}{cc}

 \label{tab:resultsSSCwitherrors}
\tablecaption{Best fit values for the SSC model parameters with estimated errors.}

\tablehead{\colhead{Parameter} & \colhead{Value} }
\decimalcolnumbers
\startdata
     Magnetic Field $B$ (G) & $ 0.046 \pm 0.003$  \\
  Doppler Factor $\delta$ & $ 4.3\pm 0.2$\\ \hline
  Electron energy distribution parameters \\ \hline
   Power law index (lower energies) $p_1$ &$1.52 \pm 0.02$\\
 Power law index (higher energies) $p_2$ & $ 3.53 \pm 0.02 $\\
    Break Lorentz factor $\gamma_c^\prime$ &$ 3.80 ^{+0.06}_{-0.05} \times 10^3$     \\ \hline
    $\chi^2_\nu$(d.o.f) & 26.1 (20) \\
    \enddata
   
\end{deluxetable*}

\begin{figure*}
\gridline{\fig{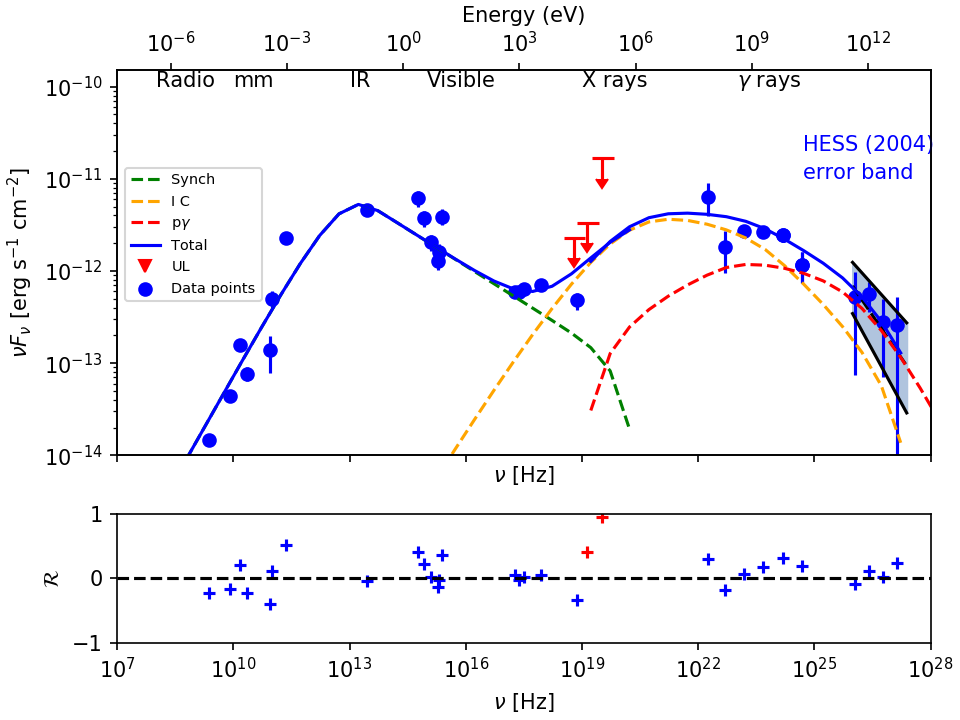}{0.4\textwidth}{(a)}\fig{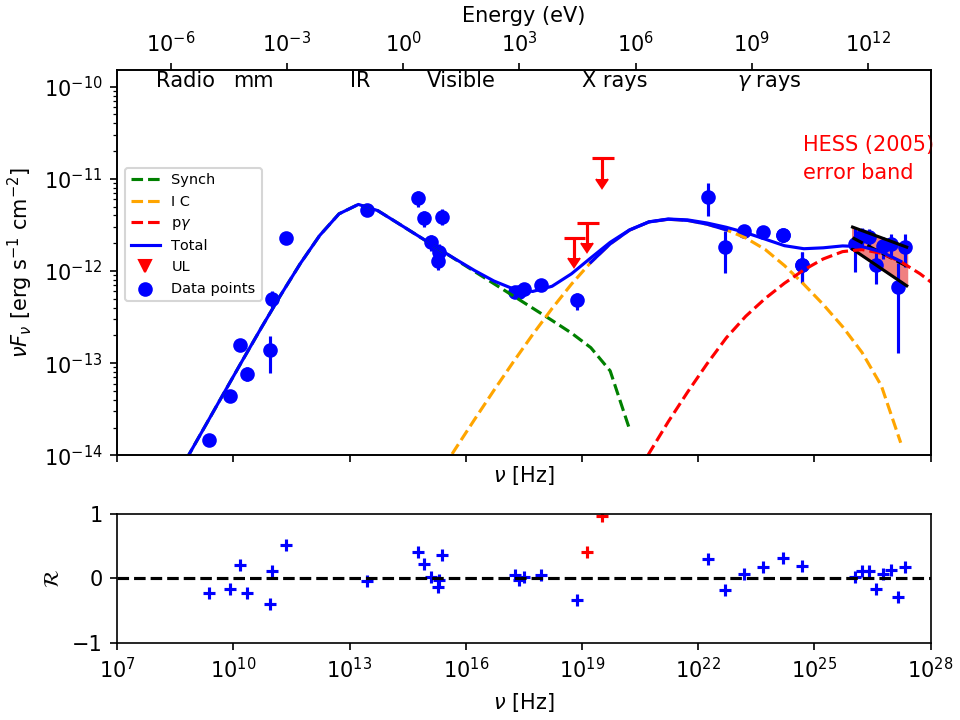}{0.4\textwidth}{(b)}}
\gridline{\fig{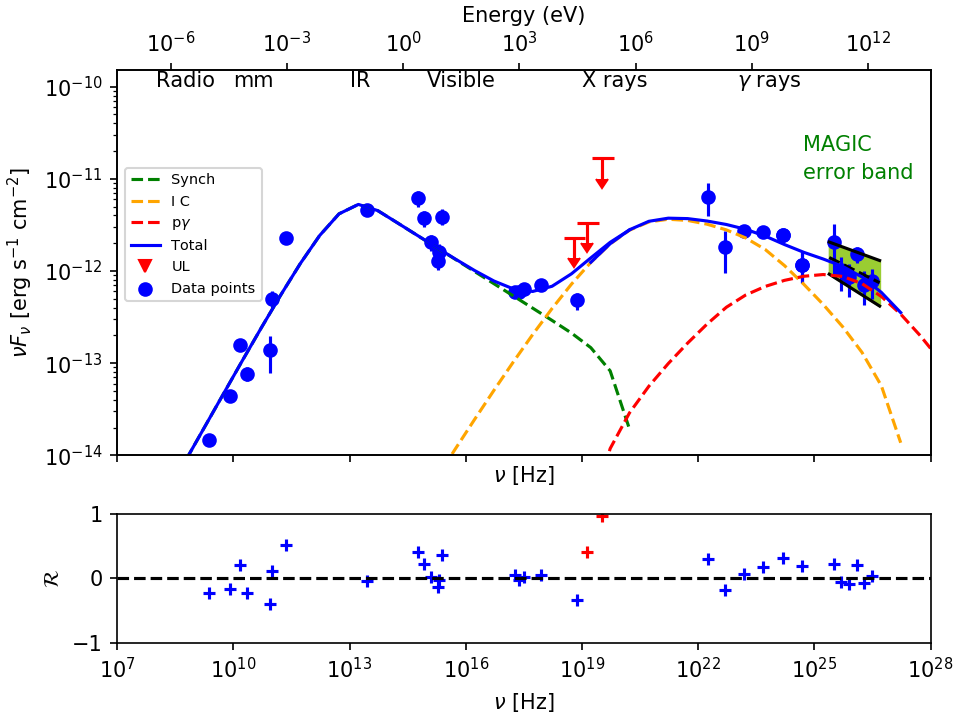}{0.4\textwidth}{(c)}\fig{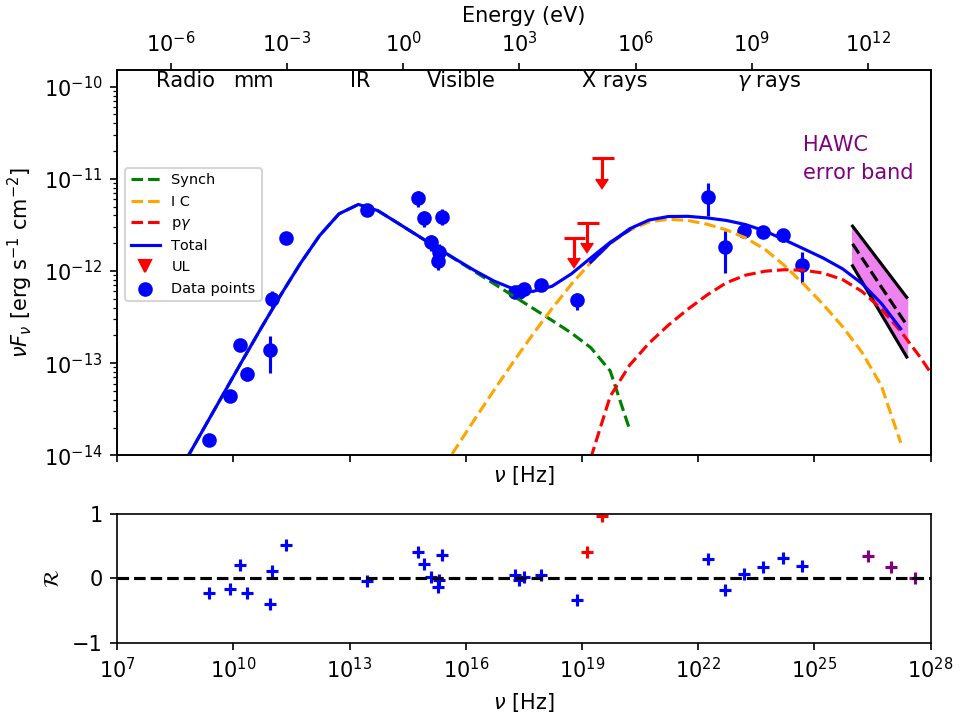}{0.4\textwidth}{(d)}}
\caption{SED of M87 with the photo-hadronic model fit for data from 2004 H.E.S.S. data \citep{Aharonian2006fast} (a), 2005 H.E.S.S. data \citep{Aharonian2006fast} (b), 2005-2007 MAGIC  data  \citep{aleksic2012magic}(c) and 2014-2019 HAWC \citep{agnsurvey}(d). The SSC model is identical to that of Figure \ref{fig:resSSC}. Measured  fluxes are plotted in blue.  The synchrotron component is  the orange dashed curve  and the inverse Compton component is the green dashed curve. \textit{Swift}/BAT upper limits are shown with red triangles. The model of the photo-hadronic component is the  red dashed curve. Residuals of the models, which are defined as $\mathcal{R}=\log(F_{\nu,obs}/F_{\nu,mod})$ where  $F_{\nu,obs}$ and $F_{\nu,mod}$ are the observed and predicted flux respectively, are shown below each plot. }
\label{fig:resph}
\end{figure*}

As mentioned before, we considered four different VHE data sets for modeling the VHE emission of M87. The best fit values for the photo-hadronic model parameters are presented in Table \ref{tab:phreserr}. The best fit models, data and residuals are plotted in Figure \ref{fig:resph}.

\begin{deluxetable*}{cccc}
\tablecaption{Best fit values for the photo-hadronic component fitting parameters with their error estimates}
\tablehead{\colhead{ } & \colhead{$\alpha$} & \colhead{$A_\gamma$} &   $\chi_\nu^2$(d.o.f)}
\label{tab:phreserr}
\decimalcolnumbers

\startdata
    H.E.S.S. :  2004 observations $^a$ & $3.2 ^{+0.2}_{-0.4} $ &  $0.1 ^{+0.2}_{-0.1} $ & 25.5 (22) \\
    H.E.S.S.: 2005 observations $^a$ & $2.8 \pm 0.2 $   &  $0.6 ^{+0.4}_{-0.2}$ &  22.5 (22) \\ 
    MAGIC: 2005-2007 observations $^b$ & $3.0 \pm 0.2$ & $0.2^{+0.2}_{-0.1}$& 23.7 (26) \\ 
    HAWC observations (1523 days) $^c$ & $3.1 \pm 0.2 $   &   $0.2\pm 0.1 $ & 25.8 (22)\\  
\enddata

\tablecomments{$\alpha$ Proton energy distribution index, $A_\gamma$ normalization constant\\
$^a$ Data taken from \cite{Aharonian2006fast}\\
$^b$ Data taken from \cite{aleksic2012magic}\\
$c$ Data taken from \cite{agnsurvey}.
}
\end{deluxetable*}

\section{Discussion}
\label{sec:discussion}

We constructed a  broadband SED  using  data from steady state epochs. It was fit with a lepto-hadronic model in order to explain its VHE emission. The lepto-hadronic model is a combination of the one-zone SSC model proposed by \cite{finke2008synchrotron} and the photo-hadronic model used by \cite{sahu2019multi2}.  The model proposes the existence of three  components: a synchrotron dominated component from radio to X-rays, an inverse Compton dominated component from X-rays to GeV gamma rays and a photo-hadronic dominated component in the range of TeV gamma rays. \\

As shown in Figure \ref{fig:resSSC}, the SSC model provides a good fit to the SED up to GeV gamma rays, but it underestimates the VHE emission,  consistently with what is found in the literature \cite[e.g.][]{fraija2016neutrino,benkhali2019complex}. The X-ray emission was of particular interest  for corresponding to the transition between both components of the SSC scenario. X rays are also important to produce the VHE emission in the photo-hadronic scenario, because the typical energies of the target photons for the photo-hadronic interactions fall in the X-ray range. The  only X-ray data point which could not be well fit was the \textit{NuSTAR} 20-40 KeV  observation. In \cite{wong2017hard}, which  reported this observation, they also discussed its inconsistency with inverse Compton models to explain gamma ray emission.  According to them, the observation uncertainties are limited by the statistical power of their data and deeper observations are necessary to resolve the tension with the NuSTAR data. \\

Table \ref{tab:paramvarios} shows a comparison of  the SSC parameters values   obtained by different studies.  Two of them (the magnetic field intensity and the break Lorentz factor) present a dispersion of several orders of magnitude. This dispersion can be caused by degeneracy in the SED models,  different  assumptions in the emission zone geometry or different level of completeness in  the  data sets. In the case of magnetic field intensity ($B$), most of the degeneracy comes from the relation  $B\delta
^2R_b^\prime\approx$ constant \citep{finke2008synchrotron}, which was first derived by \cite{tavecchio1998constraints}. As this relation holds in every case of Table \ref{tab:paramvarios} ($B\delta
^2R_b^\prime\approx (2-4)\times10^{16}$), variations in the magnetic field intensity ($B$) can be caused by different assumptions regarding the radius of the emission zone $R_b^\prime$.  The other parameters have less dispersion. It is important to remark that the lack of error estimates for most of the parameters reported in the literature prevents a more precise comparison of these results. \\

\begin{deluxetable*}{cccccc}
\tablehead{ Parameter & This study & \cite{abdo2009fermi} & \cite{de2015high}& \cite{fraija2016neutrino} & \cite{acciari2020monitoring}}
\decimalcolnumbers
\startdata
    $B$(G) &     $ 0.046 \pm 0.003$ & $0.055$ & $0.002$ & $1.61$ & $0.0031$\\ 
        $\delta$ &$ 4.3\pm 0.2$&$3.9$ &$5$  & $2.8$ & $5.3$\\
        $p_1$ &$1.52 \pm 0.02$ &$1.6 $ & $-1.8$ & $3.21\pm 0.02$ & $1.9$ \\
        $p_2$ & $ 3.53 \pm 0.02 $ & $3.6$ & $3.4$ & $4.21$ & $3.2$ \\
        $\gamma_c^{\prime}$ & $ 3.80 ^{+0.06}_{-0.05} \times 10^3$ & $4\times 10^3$ & $ 4.0\times10^2$ & $ 1.7\times10^3$ & $1.4\times10^4$\\ 
        $R_b^{\prime}$(cm)& $(1.54\pm0.07)\times10^{16}$&$1.4\times10^{16}$&$5.6\times10^{17}$&$2.1\times10^{15}$ &$4.0\times10^{17}$\\ 
\enddata
\caption{Comparison of results for the SSC parameters from different studies. All the SEDs were constructed to model the non-flaring state of M87}
\label{tab:paramvarios}
\end{deluxetable*}

The viewing angle ($\theta$) plays an important role in AGN properties. According to unification schemes, the transition between blazars and RDGs is produced around $\theta\sim10^{\circ}$.   The Doppler factor can be constrained from $\theta$ by \citep{abdo2010fermi}:
\begin{equation}
\delta\le\frac{1}{\sin(\theta)} .
\end{equation}
Higher Doppler factors enhance the HE and VHE emission, which explains why many more blazars have been detected at gamma ray energies than RDGs.\\

The Doppler factor value  obtained in this work ($\delta=4.3\pm 0.2$) is consistent with the lowest estimates for the viewing angle ($\theta\approx10^{\circ}-20^{\circ}$) \citep{biretta1999hubble}, which have been based mainly on optical observations of the jet feature \textit{HST}-1. However, it is in disagreement with other estimates  ($\theta\approx30^{\circ}-45^{\circ}$) \citep{ly2007high} based on VLBI observations of the jet base of M87. In fact, all  the  values for $\delta$ presented in Table \ref{tab:paramvarios} are  inconsistent with the VLBI measurements. One way to solve this tension is taking into account the width of the jet base. According to \cite{hada2016high} the M87 jet base has an apparent opening angle of $\sim 100^{\circ}$, which would correspond to an intrinsic opening angle of $\sim 50^{\circ}$ (if $\theta\sim 30^{\circ}$). Therefore, a VHE emission zone located in the outer zones of the jet base could have a viewing angle as low as $\sim 5 ^{\circ}$, which would be consistent with a Doppler factor $\delta\le11.8$. \\

As shown in Figure \ref{fig:resph}, the photo-hadronic component is able  to explain the VHE emission in both flaring and quiescent states. This component is produced by the interaction between inverse Compton photons and accelerated protons from the jet. We studied the quiescent state  using three different sets of TeV data: H.E.S.S. observations from 2004, MAGIC observations from 2005-2007  and HAWC observations from the 2014-2019 period (all three of them without VHE flares). The high activity state was studied  using H.E.S.S. data corresponding to the 2005 VHE flare. There were no reports of inner jet high activity in other bands during that flare. That is why we used the same broadband SED  as in the non-flaring state. HAWC  results were in agreement with the 2004 H.E.S.S.  observations and the MAGIC observations, but HAWC fluxes are lower than the fluxes of the 2005 flaring state.  This indicates that HAWC observations constrain the average VHE emission from M87 during quiescent  periods.\\

The proton energy distribution index $\alpha$ was estimated with the four VHE data sets . Those measurements agree with the  result obtained by \cite{fraija2016neutrino} $\alpha=2.80\pm0.02$ where a similar lepto-hadronic model was used to fit the 2004 H.E.S.S. data. It is also important to remark the change in $\alpha$ observed in the 2005 H.E.S.S. observation with respect to other TeV results and the lack of high activity in the other  bands during this flare. Those results indicate that the flare could have been caused by an energy increase of the accelerated proton population.  \\

It is important to mention that HAWC data are the only continuous TeV measurements. The IACT observations are made with exposure times corresponding to a few hours that could be affected by rapid VHE flux variations.  Therefore, HAWC results represent a steadier constraint to the mean VHE emission of M87 during the  HAWC's period of observation. \\

With these results, we calculated an electron luminosity of $L_e \sim 7\times10^{42}$ erg/s. Accelerated proton flux ($F_p$) and luminosity ($L_p$) can also be estimated using Equation below \citep{sahu2019multi2},

\begin{equation}
F_p(E_p)\approx 10\times \frac{f^{p\gamma}(\epsilon_\Gamma)}{\tau_{p\gamma}(E_p)},
\label{eq:proton_flux}
\end{equation}
where  $f^{p\gamma}(\epsilon_\Gamma)$ is given by Equation \ref{eq:ph} and $\tau_{p\gamma}$ is the optical depth of the $\Delta$ resonance process in the inner jet region. As mentioned before, this model assumes the photo-hadronic interactions to occur in an inner compact region of the blob with a smaller size and a higher photon density. Unfortunately, these quantities are not directly observable and the value of $\tau_{p\gamma}$ cannot be calculated. However, \cite{sahu2019multi2} gives two prescriptions  $\tau_{p\gamma}<$ 2 and $\tau_{p\gamma}>f^{p\gamma}(\epsilon_\Gamma)/f_{Edd}$, where $f_{Edd}$ is the Eddington flux. As in this case $f^{p\gamma}(\epsilon_\Gamma)/f_{Edd}\approx10^{-7}$, we assumed the intermediate value $\tau_{p\gamma}\approx10^{-2}$ and we obtained a proton luminosity of $L_p\sim6\times10^{43}$ erg/s. These results are in agreement with the total jet power estimates $L_j \sim 10^{44}$ erg/s \citep{owen2000m87}.

The decay of charged pions produces neutrinos. The neutrino flux ($f^{\nu}$) can be estimated assuming \citep{PhysRevLett.124.051103}:

\begin{equation}
    E_\nu\approx\epsilon_\Gamma/2 ,  
\end{equation}

where $E_\nu$ is the emitted neutrino energy, and also \citep{murase2016}: 
\begin{equation}
f^{\nu}(E_\nu)\approx\frac{3}{4} f^{p\gamma}(\epsilon_\Gamma/2).
\end{equation}

The estimated neutrino flux for $E_\nu=5$ TeV corresponds to $\epsilon_\Gamma=10$ TeV, is $f^{\nu}\sim 1\times10^{-13}$ TeV cm$^{-2}$ s$^{-1}$, which is below IceCube upper limits \citep{PhysRevLett.124.051103}.

As with regard to the results of some other alternative models,  in \cite{fraija2016neutrino} the SED of M87 was fit with a very similar lepto-hadronic model. In this case, VHE emission was  represented only by the 2004 H.E.S.S. results, which were obtained with just $\sim50$ hours of observation. However, the best fitting values of the photo-hadronic  parameters were in agreement, within their uncertainties,  with those obtained in this work. In \cite{sahu2015hadronic} a lepto-hadronic scenario was used to explain the 2010 TeV flare and their results were also in agreement with those obtained in this work. However, this flare had an X-ray counterpart \citep{abramowski20122010}, which may be interpreted as purely leptonic. A leptonic and a hybrid model were used to model MAGIC results from a 2012-2015 campaign \citep{acciari2020monitoring}, where they found the hybrid model was more consistent with gamma ray data. Finally, according to  \cite{benkhali2019complex}, extended gamma ray production scenarios such as Compton scattering in the kiloparsec-scale jet \citep[e.g.,][]{hardcastle2011modelling} are disfavored by gamma ray variability.  \\

 We cannot rule a purely leptonic scenario out. Actually, multi-zone structured leptonic models are necessary to explain specific features in the whole MWL SED \citep{algaba2021broadband}. However, these models have a large number of parameters, which introduces a high degeneracy, making it difficult to derive firm conclusions from their fit. As the aim of this paper is explaining the VHE emission, we decided to use a one-zone SSC scenario to model the leptonic contribution

\section{Conclusions}

M87 is considered a laboratory for understanding AGN properties since it is the only source that has been mapped from the SMBH shadow ($\sim 0.005$ pc) to the outer jet ($\sim 25$ kpc). Understanding its gamma ray emission, which is practically unaffected by EBL absorption up to $\sim$ 10 TeV, could be the key to understanding gamma ray emission in the rest of the AGNs (not only RDGs). We fit a broadband SED of M87 with a lepto-hadronic model with the aim of explaining its VHE emission. Emission from radio to GeV gamma rays has been modeled with an SSC scenario. The best fit values for SSC model parameters were for the  Doppler factor $\delta=$ $4.3\pm 0.2$, for the mean magnetic field intensity $B=$ $ 0.046 \pm 0.003$ G, for the electron energy distribution parameters  $p_1=$ $1.52 \pm 0.02$, $p_2=$ $ 3.53 \pm 0.02  $ and $\gamma_c^{\prime}=$ $ 3.80 ^{+0.6}_{-0.5} \times10^{3}$. The value of the Doppler factor is in agreement with a low viewing angle of the jet  base ($\theta\sim 13
^\circ$). However, a large viewing angle is also possible if the opening angle of the jet base is wide enough to place the emission zone closer to the observer's line of sight.\\

A photo-hadronic model was fit to the VHE emission. Results show that this model is able to explain the quiescent VHE  emission represented by H.E.S.S., MAGIC and HAWC observations. H.E.S.S. data corresponding to the 2005 VHE flare were also fit using this model. The results show that the  model can explain the so called orphan flares, which are only detected at VHE bands, such as the one observed in 2005. Those flares would be produced by changes in the proton energy distribution.\\

HAWC observations constrained the VHE emission from  M87 for the 2014-2019 period in which no evidence of VHE flares was reported. 
We obtained a proton energy distribution power law index of $\alpha=3.1 \pm 0.2$ and TeV gamma ray flux normalization constant of $A_\gamma=0.2\pm0.1$. 
HAWC will be  taking  data for a few more years. Therefore, the significance of the M87 detection will probably be improved allowing  a better estimation of the photo-hadronic model parameters.\\

\section{ACKNOWLEDGMENTS}
We acknowledge the support from: the US National Science Foundation (NSF); the US Department of Energy Office of High-Energy Physics; the Laboratory Directed Research and Development (LDRD) program of Los Alamos National Laboratory; Consejo Nacional de Ciencia y Tecnolog\'ia (CONACyT), M\'exico, grants 271051, 232656, 260378, 179588, 254964, 258865, 243290, 132197, A1-S-46288, A1-S-22784, c\'atedras 873, 1563, 341, 323, Red HAWC, M\'exico; DGAPA-UNAM grants IG101320, IN111716-3, IN111419, IA102019, IN110621, IN110521; VIEP-BUAP; PIFI 2012, 2013, PROFOCIE 2014, 2015; the University of Wisconsin Alumni Research Foundation; the Institute of Geophysics, Planetary Physics, and Signatures at Los Alamos National Laboratory; Polish Science Centre grant, DEC-2017/27/B/ST9/02272; Coordinaci\'on de la Investigaci\'on Cient\'ifica de la Universidad Michoacana; Royal Society - Newton Advanced Fellowship 180385; Generalitat Valenciana, grant CIDEGENT/2018/034; Chulalongkorn University’s CUniverse (CUAASC) grant; Coordinaci\'on General Acad\'emica e Innovaci\'on (CGAI-UdeG), PRODEP-SEP UDG-CA-499; Institute of Cosmic Ray Research (ICRR), University of Tokyo, H.F. acknowledges support by NASA under award number 80GSFC21M0002. We also acknowledge the significant contributions over many years of Stefan Westerhoff, Gaurang Yodh and Arnulfo Zepeda Dominguez, all deceased members of the HAWC collaboration. Thanks to Scott Delay, Luciano D\'iaz and Eduardo Murrieta for technical support.

\software{Python \citep{van1995python}, NumPy \citep{harris2020array}, Astropy \citep{astropy:2018}, Matplotlib \citep{Hunter:2007}}

\bibliography{biblio}
\bibliographystyle{aasjournal}



\end{document}